\documentclass[conference]{IEEEtran}
\usepackage[noadjust]{cite}
\usepackage{amsmath,amssymb,amsfonts}
\usepackage{algorithmic}
\usepackage{graphicx}
\usepackage{textcomp}
\usepackage[T1]{fontenc}

\ifCLASSINFOpdf
\else
\fi
\begin{document}
%
\title{Hardware Implementation of A Non-RLL Soft-decoding Beacon-based Visible Light Communication Receiver}




%
\author{\IEEEauthorblockN{Duc-Phuc Nguyen\IEEEauthorrefmark{2},
Dinh-Dung Le\IEEEauthorrefmark{2},
Thi-Hong Tran\IEEEauthorrefmark{2}, 
Huu-Thuan Huynh\IEEEauthorrefmark{3} and
Yasuhiko Nakashima\IEEEauthorrefmark{2}}
\IEEEauthorblockA{\IEEEauthorrefmark{2}Graduate School of Science and Technology, Nara Institute of Science and Technology (NAIST)\\
630-0192 Takayama-cho, Ikoma-shi, Nara-ken, Japan\\ Email: (nguyen.phuc.ni6, le.dung.ku9, hong, nakashim)@is.naist.jp}
\IEEEauthorblockA{\IEEEauthorrefmark{3}Ho Chi Minh City - University of Science\\ 227 Nguyen Van Cu Str., Ward. 4, Dist.5, Ho Chi Minh City - Vietnam,
Email: hhthuan@hcmus.edu.vn}}


\maketitle

\begin{abstract}
Visible light communication (VLC)-based beacon systems, which usually transmit identification (ID) information in small-size data frames are applied widely in indoor localization applications. There is one fact that flicker of LED light should be avoid in any VLC systems. Current flicker mitigation solutions based on run-length limited (RLL) codes suffer from reduced code rates, or are limited to hard-decoding forward error correction (FEC) decoders. Recently, soft-decoding techniques of RLL-codes are proposed to support soft-decoding FEC algorithms, but they contain potentials of high-complexity and time-consuming computations. Fortunately, non-RLL direct current (DC)-balance solutions can overcome the drawbacks of RLL-based algorithms, however, they meet some difficulties in system latency or inferior error-correction performances. Recently, non-RLL flicker mitigation solution based on Polar code has proved to be an optimal approach due to its natural equal probabilities of short runs of 1's and 0's with high error-correction performance. However, we found that this solution can only maintain the DC balance only when the data frame length is sufficiently long. Accordingly, short beacon-based data frames might still be a big challenge for flicker mitigation in such non-RLL cases. In this paper, we introduce a flicker mitigation solution designed for VLC-based beacon systems that combines a simple pre-scrambler with a Polar encoder which has a codeword smaller than the previous work 8 times. We also propose a hardware architecture for the proposed compact non-RLL VLC receiver for the first time. Also, a 3-bit soft-decision filter is introduce to enable soft-decoding of Polar decoder to improve the performance of the receiver.  
\end{abstract}

%
\IEEEpeerreviewmaketitle

\section{Introduction}
VLC simultaneously provides both illumination and communication services. VLC system conveys modulated digital information via a transmit (TX) front-end and light-emitting diode (LED)'s light. The brightness and stability of the light are affected by the distribution of the 1\char`\'s and 0\char`\'s in the data frames. As a result, flicker mitigation which based on DC-balance techniques is considered as one of essential concerns in any VLC systems. In VLC-beacon-based indoor localization systems, unique ID information are transmitted from VLC-LED bulbs for purposes such as identifying objects and locations \cite{Yoshikawa}. Furthermore, beacon-based frames have been introduced in some publications with the sizes of 158-bit \cite{Yoshikawa}, 56-bit \cite{Qing} or 34 symbols (0.96ms) \cite{Yamazato}. We found that the 158-bit beacon-based frame which is defined by Standard of Japan Electronics and Information Technology Industries Association (JEITA) should be considered in this paper \cite{Latif,Yamazato}. 

Table \ref{table_1} summarizes proposals related to FEC and flicker mitigation for VLC. The conventional solution is defined in the IEEE 802.15.7 VLC standard and employs Reed-Solomon (RS) codes, Convolutional Codes (CC) and hard-decoding RLL (hard-RLL) algorithms \cite{Sridhar}. However, the applicability of hard-RLL methods is limited to hard-decoding FEC codes \cite{Sridhar,Wang1,Sunghwan}; consequently, the error-correction performance of the entire system is restricted. Recently, soft-decoding RLL (soft-RLL) solutions have been proposed in \cite{Wang2,Wang3,Wang4,Dung}. These techniques permit soft-decoding FEC algorithms to be applied to improve the bit-error-rate (BER) performance of VLC systems, but they also include heavy computational efforts, with many additions and multiplications. Zunaira \emph{et al.} have proposed replacing the classic RLL codes with a recursive Unity-Rate Code (URC) as the inner code and a 17-subcode IRregular Convolutional Code (IRCC) as the outer code \cite{Babar}. Although this method can achieve a dimming level of approximately 50\% with good BER performance, however, the system latency is increased with the iterative-decoding IRCC-URC scheme. In addition, the long codeword lengths, which ranges from 1000 to 5000 bits, reduce the compatibility of this proposal to VLC-based beacon systems in which beacon-based frame sizes are always small \cite{Yoshikawa,Qing} . As an alternative approach, Kim \emph{et al.} have proposed two coding schemes based on the modified Reed-Muller (RM) codes \cite{Sunghwan}. Although this method can guarantee DC balance at exactly 50\%, but it has inherent drawbacks of a low code rate and an inferior error-correction performance compared with turbo codes, low-density parity-check (LDPC) codes or polar codes. In addition, Lee and Kwon have proposed the use of puncturing and pseudo-noise sequence scrambling with compensation symbols (CSs) \cite{Lee}. This proposal can achieve very good BER performance; however, puncturing with CSs will lead to redundant bits in the messages, thereby reducing the transmission efficiency. Another coding scheme based on fountain code, with greatly improved transmission efficiency, was mentioned in \cite{Lifang}. However, this scheme requires feedback information and thus is not suitable for broadcasting scenarios in VLC-based beacon systems. Recently, Xuanxuan Lu \emph{et al.} have reported a new class of enhanced Miller codes, termed eMiller codes which is a class of RLL codes known for high-bandwidth efficiency \cite{Xuanxuan}. She also proposed an improved version of Viterbi algorithm, termed \emph{mnVA} to further enhance the performance of her proposed eMiller. It can be seen from her simulation results that eMiller helps improve the performance of the whole VLC system; and this code seems to be a promising for VLC applications. However, two main drawbacks of this approach are the code-rate = 1/2 is still not optimized if compared with non-RLL approaches, and an increasing in computational complexity.      

\begin{table}[!t]
\renewcommand{\arraystretch}{1.3}
\caption{Overview of FEC and flicker mitigation solutions for VLC}
\label{table_1}
\centering
\begin{tabular}{c | c}
\hline
\bfseries FEC solution & \bfseries Flicker mitigation\\
\hline \hline
RS, CC \cite{Sridhar} & Hard-RLL\\\hline
Multi-RS hard-decoding \cite{Wang1} & Hard-RLL\\\hline
LDPC \cite{Kim} & Hard-RLL \\\hline
RS soft-decoding \cite{Wang2,Wang3} &Soft-RLL \\\hline
Polar code \cite{Wang4,Dung} & Soft-RLL \\\hline
Irregular CC \cite{Babar}& Unity-Rate Code\\\hline
Reed-Muller \cite{Sunghwan}& Modified original code\\\hline
Turbo code \cite{Lee} & Puncture + Scrambling\\\hline
Fountain code \cite{Lifang} & Scrambling\\\hline
Convolutional code, Viterbi \cite{Xuanxuan} & Enhanced Miller code\\\hline
Polar code ($N$=2048) \cite{Fang}& Flicker-free\\\hline
Pre-scrambled polar code\\ 
(JEITA's beacon frame size),\\
proposed method ($K$=158, $N$=256 )& Flicker-free\\\hline
\end{tabular}
\end{table}

Advantages of Polar code are exploited deeply together with soft-decoding of RLL codes have been introduced at \cite{Wang4,Dung}. According to these publications, Manchester and 4B6B codes are used as RLL solutions for the VLC transmitter. As a result, their BER performances have been improved remarkably with a flexibility of Polar code's code-rate. However, we found that code-rate = 1/2 of Manchester code, or code-rate = 0.67 of 4B6B are not the best optimization solution for transmission efficiency enhancement, if compared with non-RLL approaches. Moreover, soft-decoding of RLL codes are time-consuming works which includes multiplication, exponential and logarithm computations. 
Fang \emph{et al.} have recently proposed a non-RLL polar-code-based solution for dimmable VLC \cite{Fang}. This approach has shown promising results in weight distribution and run-length distribution. Moreover, this solution also show an improved transmission efficiency while achieving a high coding gain compared with RS and LDPC codes. We have found that this solution can overcome the drawbacks of related works mentioned above; specifically, it offers non-iterative decoding, a flexible code rate, and a high BER performance without requiring feedback information. However, we also found that the big obstacle of this proposal is the equal probabilities of short runs of 1\char`\'s and 0\char`\'s can only be achieved with a long codeword length; as chosen to be \textit{N}=2048. Although, it is stated that, in low-throughput VLC systems, such as beacon-based ones, long data frames must be avoided. It is evident that a solution \cite{Fang} based on a polar encoder alone is not applicable in such VLC-based beacon systems because DC balance is not guaranteed for short data frames. 
In this paper, we propose a fast-convergence non-RLL DC-balance scheme based on a pre-scrambled polar encoder. Consequently, the proposed method can guarantee DC balance around the range of (41.25\%, 63.75\%) by a Polar encoder has a codeword length of \textit{N}=256 and input frame length equals 158-bit (JEITA's beacon-frame length) \cite{Latif,Yamazato}, which is 8 times shorter than the codeword length of \textit{N}=2048 required for the polar-code-based solution proposed in \cite{Fang}. Furthermore, as an advantage of the non-RLL solution, our proposal also show a better code-rate (0.617) compared with RLL related works mentioned in Table.\ref{table_1}.

\section{Our Proposed System}
\begin{figure}[!t]
\centerline{\includegraphics[width=\columnwidth]{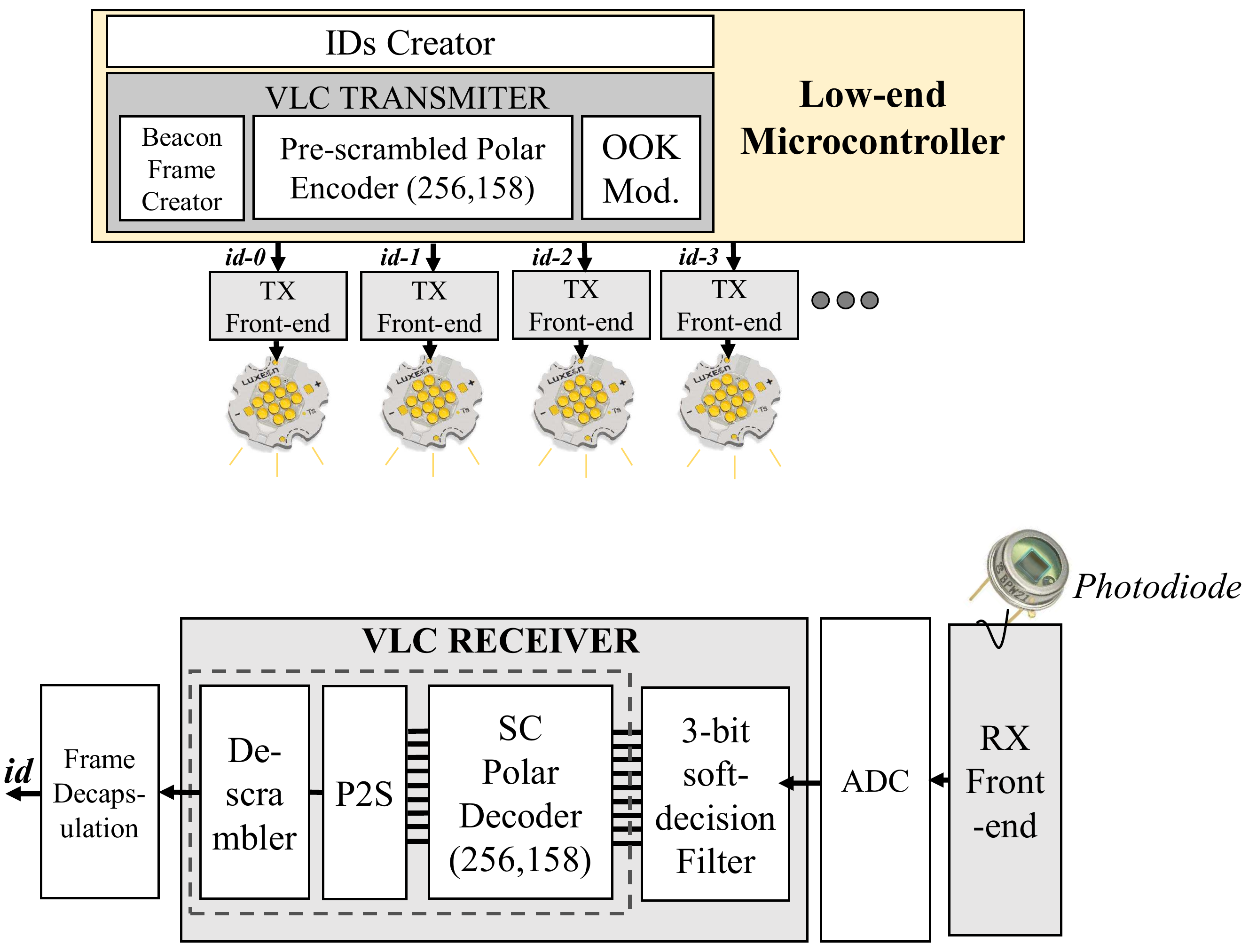}}
\caption{Block diagram of our beacon-based VLC receiver}
\label{fig1}
\end{figure}
Our proposed system is presented in Fig.\ref{fig1} in which VLC transmitter's functions are processed by a firmware program on a low-end micro-controller; while the VLC receiver is implemented on an FPGA with a novel hardware architecture.

In a digital transmission system, a data scrambler plays an important role because it causes energy to be spread more uniformly. At the transmitter, a pseudorandom cipher sequence is modulo-2 added to the data sequence to produce a scrambled data sequence. 

Describe the generating polynomial P(x) by Eq.\ref{eq:polynomial}.
\begin{equation}\label{eq:polynomial}
P(x)=\displaystyle\sum_{q=0}^{N} c_q.x^q
\end{equation}
where $c_0$ = 1 and equals 0 or 1 for other indexes.

We have found that the output bit probability distributions (BPD) of the pre-scrambler in different generating polynomials seem to differ slightly. Therefore, we propose a simple generating polynomial presented in Eq.(\ref{eq:scrambler}): 
\begin{equation}\label{eq:scrambler}
P(x)=x^{15} + x^{14} + 1
\end{equation}
 
Meanwhile, Polar codes can be classified into two types: non-systematic and systematic codes. Typically, a polar code is specified by a triple consisting of three parameters: \textit{(N, K, I)}, where \textit{N} is the code length, \textit{K} is the message length, and \textit{I} is the set of information bit indices. Let \textit{d} be a vector of \textit{N} bits, including information bits. The generator matrix is defined as \(G=(F^{\otimes n})_I\). Then, given a pre-scrambled message \textit{u} of \textit{K} bits in length, a codeword \textit{x} is generated as Eq.(\ref{eq:polarencode}). 
\begin{equation}\label{eq:polarencode}
x = u.G = d.F^{\otimes n}
\end{equation}
Systematic polar codes were introduced to achieve a better error-correction performance compared with non-systematic polar codes \cite{Harish}. A Polar encoder is formed by many layers of XOR gates, with a complexity of \(\frac{N}{2}log_2N\) XORs. The output bit probability distribution of a Polar encoder naturally becomes centralized at approximately 50\% 1\char`\'s and 50\% 0\char`\'s when the codeword length increases \cite{Fang}. 
We have selected the Polar code as the main FEC scheme for our non-RLL VLC transmitter/receiver for several reasons:
\begin{enumerate}
\item Unusual code rates are supported. Specifically, a (256;158) polar code, which has a code rate of 0.617, is suitable for a beacon-based frame size defined by JEITA \textit{K}=158 \cite{Yoshikawa,Latif}.
\item The encoder's output bit probability distribution is naturally centralized at 50\% 1\char`\'s and 50\% 0\char`\'s when codelength is long enough\cite{Fang}.
\item High error-correction performance can be achieved with low hardware complexity \cite{Phuc}.
\item The inherently short run lengths of a polar encoder can help mitigate the lighting flicker \cite{Fang}.
\end{enumerate}

Through experiment results, we have found that a pre-scrambler can help to ensure a fast convergence of the output probability distribution of an inner (256;158) Polar encoder. As a result, DC balance in a VLC-based beacon system can be guaranteed by the proposed system depicted in Fig.\ref{fig1}. On-Off Keying (OOK) modulation is considered because of its simplicity. Also, at the receiver, we have implemented a soft-decision filter which extracts log-likelihood ratio (LLR) values calculated from received signals' voltages.
Specifically, in the case of VLC AWGN channel, the \textit{log-likelihood ratio} (LLR) values could be calculated by Eq.(\ref{eq:llr}). 
\begin{equation} \label{eq:llr}
LLR(y_i) = \ln{\frac{P(x_i=0|y_i)}{P(x_i=1|y_i)}}
\end{equation}
where $y_i$ is the received sample and the conditional probability is generally calculated by Eq.(\ref{eq:probability}).
\begin{equation}  \label{eq:probability}          
P(x_i|y_i = \Delta) = \frac{1}{\sqrt{2\pi\sigma_{\Delta}^{2}}}e^{-\frac{(y_i-\mu_{\Delta})^2}{2\sigma_{\Delta}^{2}}}
\end{equation}  

where $\mu_{\Delta}$ and $\sigma_{\Delta}$ are the mean value and standard deviation for $\Delta = 0, 1$. However, when making real prototype of soft-decoding VLC receiver, we found that it is unfeasible in estimating the LLRs using such Eq.(\ref{eq:llr}) and Eq.(\ref{eq:probability}) due to $\mu_{\Delta}$ and $\sigma_{\Delta}$ can not be estimated in real wireless optical channels. Therefore, in this paper, we propose applying a soft-decision filter which is first introduced in optical communication systems for our prototype of VLC receiver \cite{Tagami}. The analog to digital converter (ADC) converts received analog signals to digital signals. The filter analyses these digital signals and calculate log-likehood ratio (LLR) values to feed to soft-decoding Polar decoder. The soft-decision filter includes $2^{N-1}$ decision thresholds to compare with the incoming received signal, where N is the number of quantization bits. Previous research on soft-decision filter in optical communication systems has shown that 3-bit soft decision was the optimum solution \cite{Tagami}. In the case of N=3 for 3-bit soft decision, seven thresholds from $V_{t+3}$ to $V_{t-3}$ are established from the error probabilities of the two possible received signals 0 and 1. We have defined a mapping table with output LLR values are carefully chosen from training simulation results on MATLAB. Table \ref{table_2} shows ranges of comparison and its output LLR values.
\begin{table}[!t]
\renewcommand{\arraystretch}{1.3}
\caption{3-bit soft-decision filter's mapping table}
\label{table_2}
\centering
\begin{tabular}{c | c | c}
\bfseries Comparator &\bfseries Range &\bfseries Output LLR values \\
\hline
\hline
0	&[$V_{peak-}$ ;  $V_{t-3}$]   &  -1.1943    \\\hline
1	&[$V_{t-3}$ ;  $V_{t-2}$]     &  -0.3547    \\\hline
2	&[$V_{t-2}$ ;  $V_{t-1}$]     &  -0.2116    \\\hline
3	&[$V_{t-1}$   ;  $V_{t}$]     &  -0.0702    \\\hline
4	&[$V_{t}$ ;  $V_{t+1}$]       &   0.0656    \\\hline    
5	&[$V_{t+1}$ ;  $V_{t+2}$]     &   0.2185    \\\hline  
6	&[$V_{t+2}$ ;  $V_{t+3}$]	   &  0.3630    \\\hline 
7	&[$V_{t+3}$ ; $V_{peak+}$]	   &  1.2017    \\\hline 

\end{tabular}
\end{table}

\begin{figure}[!t]
\centerline{\includegraphics[width=\columnwidth]{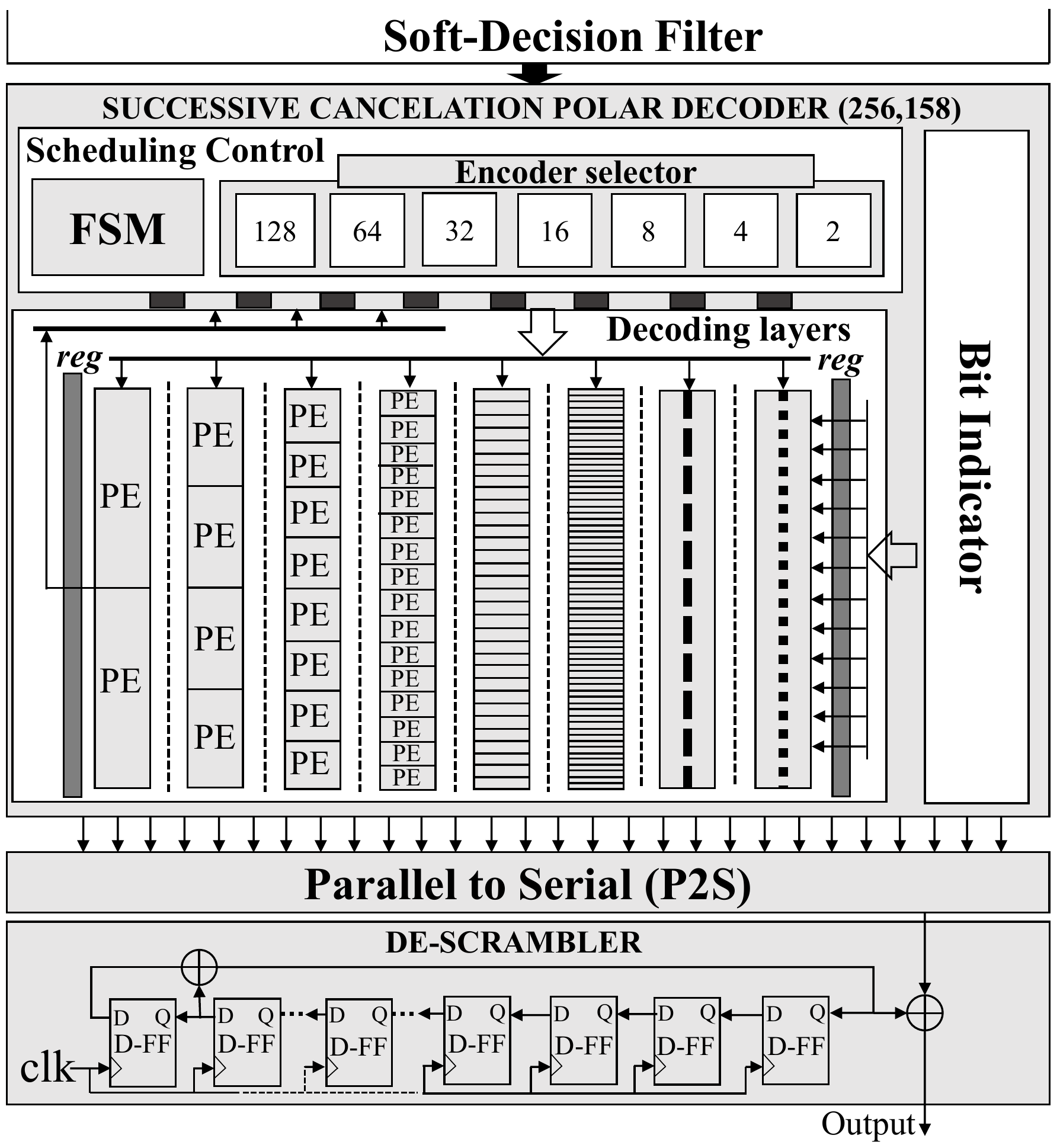}}
\caption{The proposed hardware architecture of non-RLL VLC receiver }
\label{fig2}
\end{figure}

\begin{figure}[!t]
\centerline{\includegraphics[width=\columnwidth]{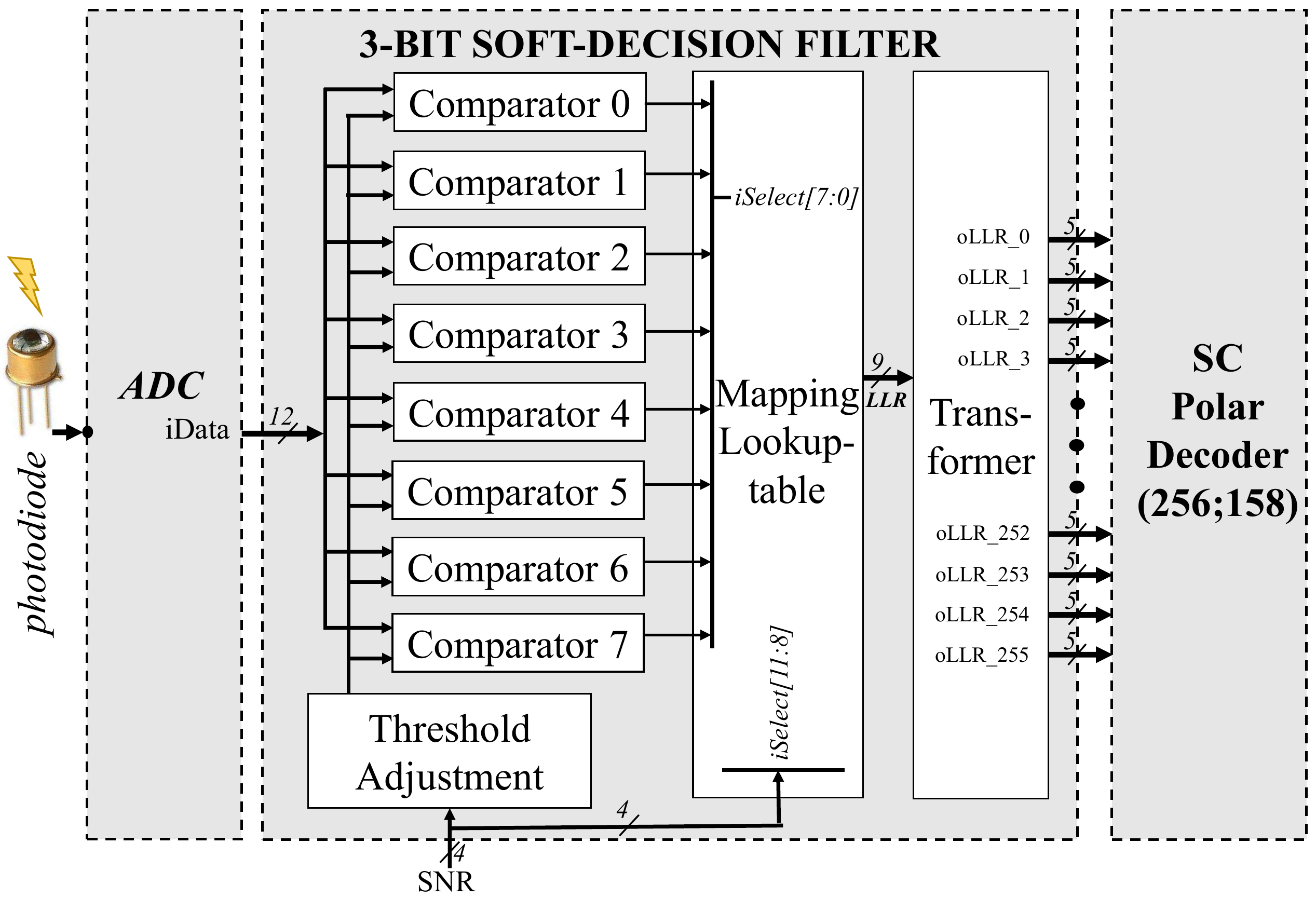}}
\caption{The proposed hardware architecture of 3-bit soft-decision filter}
\label{fig3}
\end{figure}

\section{Experimental Results}
In this section, we present evaluations of bit probability distribution, run-length, BER/FER performances on our proposal. Due to the hardware implementation, we also present the FPGA and ASIC synthesis results.
\subsection{Centralized bit probability distribution}
\begin{figure}[!t]
\centerline{\includegraphics[width=\columnwidth]{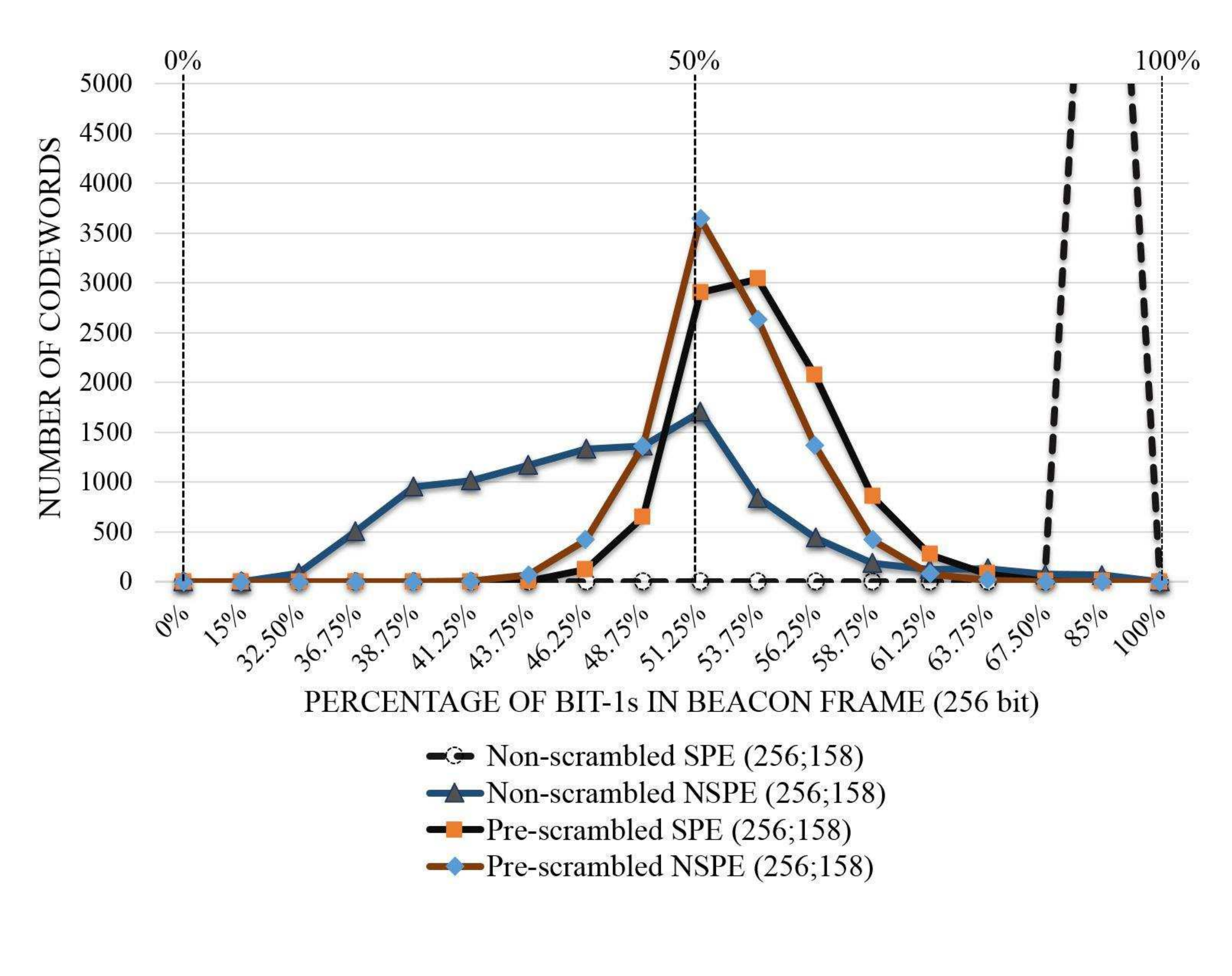}}
\caption{Bit-1 output probability distribution in beacon frames (256-bit) in cases of non-scrambled/scrambled NSPE, SPE}
\label{fig4}
\end{figure}

Regarding the previously proposed non-RLL solution based only on a polar encoder \cite{Fang}, the authors demonstrated the fluctuation of the code weight distribution around the 50\% dimming level. Specifically, in case of a polar encoder has N=2048, the percentage of 1's was reported to fluctuate in the range of (42.1875\%, 57.8125\%) \cite{Fang}. However, we have found that this fluctuation range can only be achieved when the proportions of 1\char`\'s and 0\char`\'s in the input data (before the Polar encoder) are approximately equal 50\%. Unfortunately, the bit ratio of the input data is unknown beforehand because of the randomness of the data, and this input bit ratio greatly affects the output bit ratio of the FEC encoder. If the minimum and maximum bit ratios are included, the real fluctuation range of the output bit probability distribution of a polar encoder with 2048-bit codewords is (41.25\%, 61.25\%). In this paper, we evaluate our proposed method using a bad case of input bit ratio which includes 10\% of zero bits and 90\% of one bits. A simulation was conducted with 10,000 158-bit data frames. Both systematic polar encoder (SPE) and non-systematic polar encoder are implemented to evaluate the pre-scrambler. Regarding with experimental results presented in Fig.\ref{fig4}, we can also see that an NSPE shows a better centralized bit distribution compared with that of a SPE regardless of whether pre-scrambling is applied or not. This can be explained is because the information bits transparently appear as part of the codeword, the output probability distribution of a SPE is not well centralized if compared with NSPE's one. Especially when a pre-scrambler is not used, the probability distribution of the SPE tends toward 85\% one bits which might affects strongly on flicker phenomenon. Fig.\ref{fig4} shows the impact of a pre-scrambler on the output bit probability distribution of the NSPE and SPE. Notably, DC balance is not guaranteed in the case of SPE-NSPE(256;158) if a pre-scrambler is not applied because the encoder's output bit probability distribution spreads over a large range of percentages (32.5\%, 85\%) in case of NSPE; or (67.5\%,100\%) in case of SPE. However, when a pre-scrambler is applied, the fluctuation ranges of the pre-scrambled (256;158) SPE, NSPE converge to (41.25\%, 63.75\%), whereas the fluctuation ranges of polar encoders with codeword lengths of 2048 and 1024 are (41.25\%, 61.25\%) and (38.75\%, 67.5\%), respectively. Thus, pre-scrambling causes the output bit probability distribution of a (256;158) Polar encoder to be approximately equal to those of (1024;512) and (2048;1024) encoders. In other words, considering the frame size of beacon-based systems, a pre-scrambler is necessary to ensure faster convergence to a centralized bit probability distribution. Therefore, compared with the non-RLL DC-balance solution based only on a polar encoder with 2048-bit codewords presented in \cite{Fang}, our proposed method can achieve the same output bit probability distribution with a shorter codeword length by a factor of 8. 

\begin{figure}[!t]
\centerline{\includegraphics[width=\columnwidth]{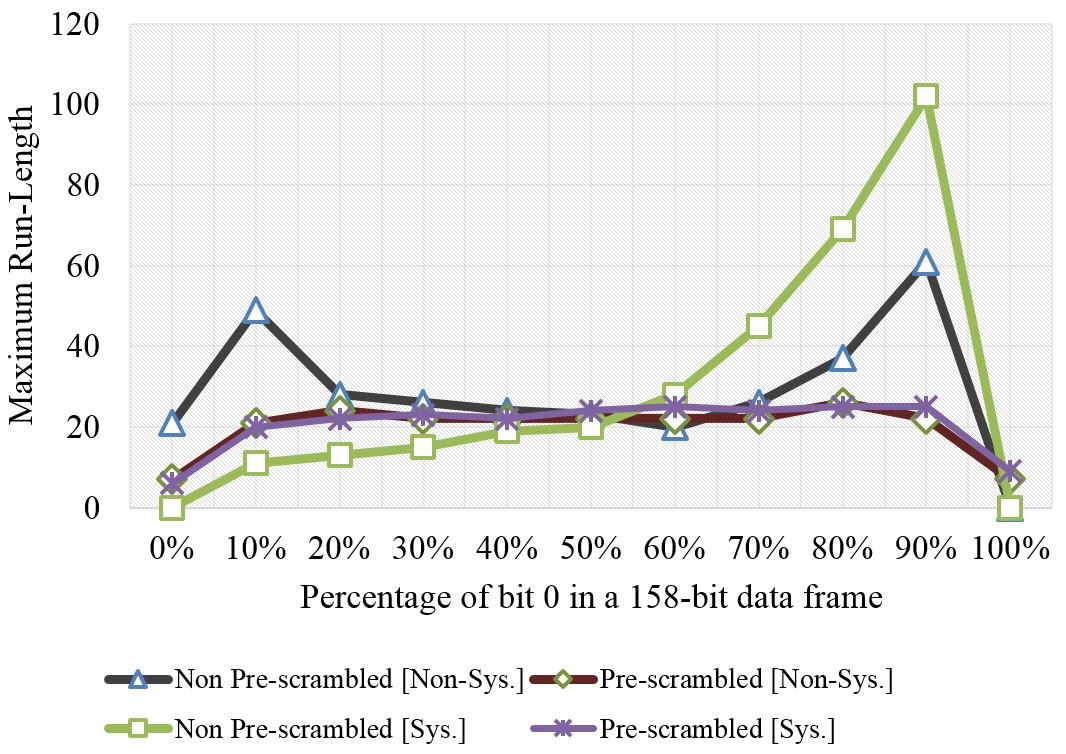}}
\caption{Run-length reduction performance when the pre-scrambler is applied for NSPE, SPE}
\label{fig5}
\end{figure}

\subsection{Run-length reduction performance}
Evaluating the flicker mitigation also requires a consideration on run-length of all frames. In the simulation model, we have sent 10000 frames of 158-bit with the percentage of bit-0 changes from 0\% to 100\%. The results have been presented in Fig.\ref{fig5}. Accordingly, the maximum run-lengths are reduced remarkably in case of the pre-scrambler is applied for Polar encoders. Specifically, the maximum run-length reduction gain that NSPE can achieve is 1.9; while an even better effect on SPE is the gain of 4.08 when 90\% of bit-0 appears in a data frame. The relationship of maximum run-length and the transmit frequency that flicker mitigation is guaranteed can be expressed by Eq.(\ref{eq:minfrequency}).        
While $F_{min-FM}$ is the minimum frequency that flicker mitigation is guaranteed; \textit{maxRL} is the maximum run-length and \textit{MFTP} is stated around 5 ms \cite{Fang}. Hence, the minimum frequency that the flicker mitigation in our proposed system is guaranteed is 2.5 Khz, which is still much smaller than the minimum frequency defined in \cite{Sridhar}. 

\begin{equation}\label{eq:minfrequency}
F_{min-FM} = \frac{1}{\text{$MFTP$ * $maxRL$}}
\end{equation}

\subsection{Bit-error-rate (BER), frame-error-rate (FER) performance}
\begin{figure}[!t]
\centerline{\includegraphics[width=\columnwidth]{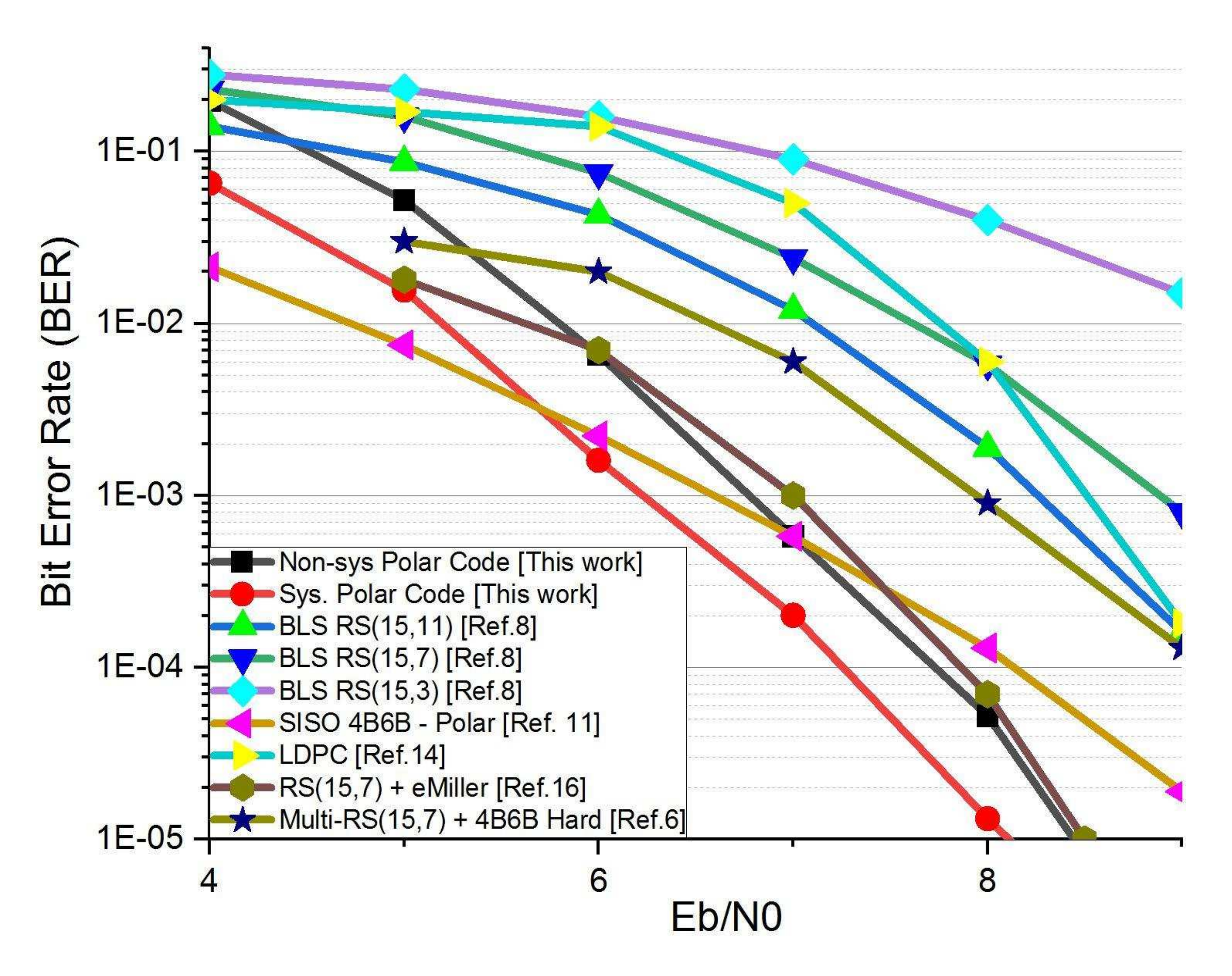}}
\caption{BER performances of the proposed VLC receiver and some related works in VLC-AWGN channel}
\label{fig6}
\end{figure}

\begin{figure}[!t]
\centerline{\includegraphics[width=\columnwidth]{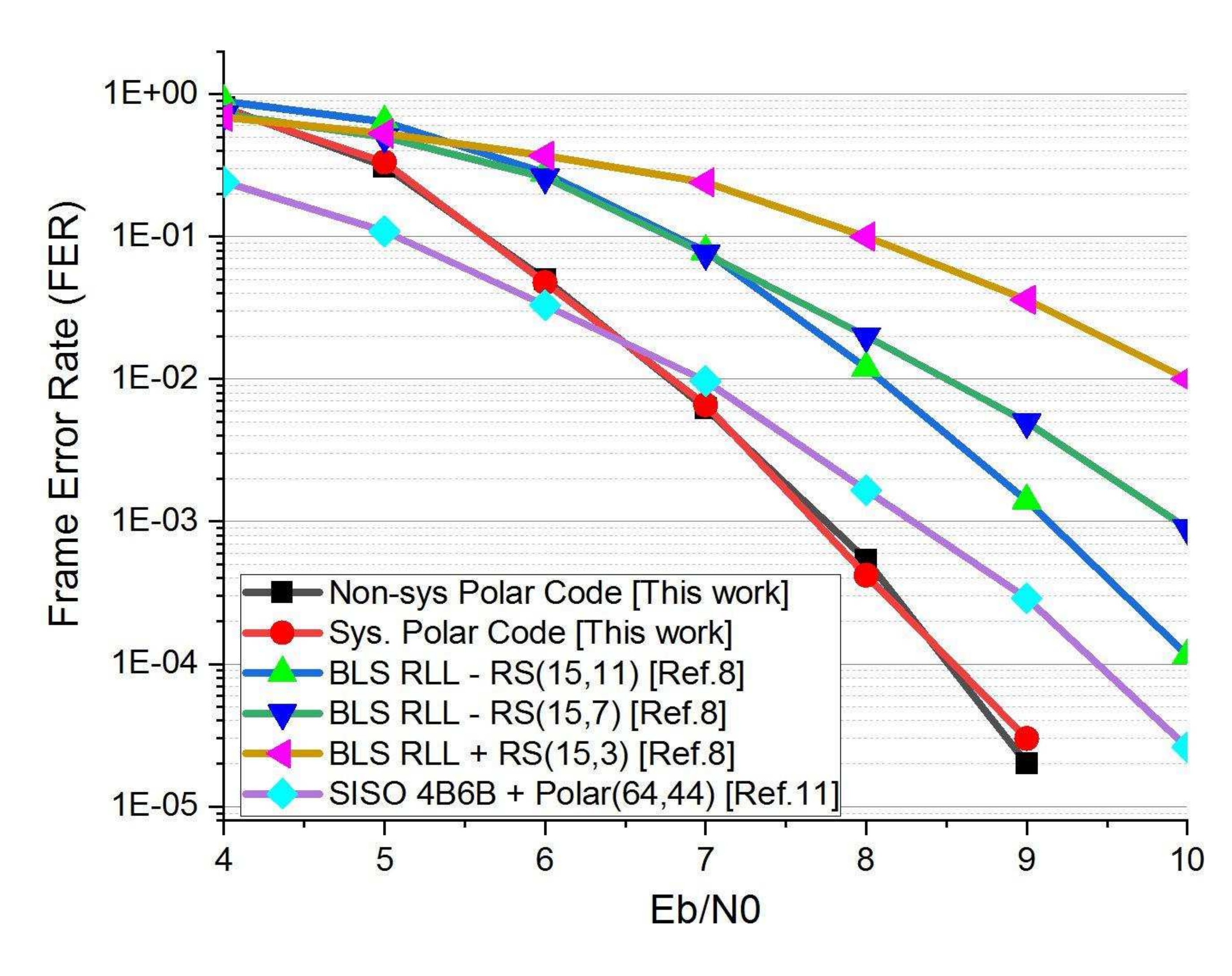}}
\caption{FER performances of the proposed VLC receiver and some related works in VLC-AWGN channel}
\label{fig7}
\end{figure}

Fig.\ref{fig6} shows the BER performances of our work compared with some related works. We both applied the non-systematic and systematic Polar codes which have code-rate = 0.617 (256;158) to evaluate the performance. Also, Fig.\ref{fig7} presents the FER performances of our works. Although systematic Polar decoder achieves better BER performance than non-systematic decoder does. However, FER performances of them are the same for all cases. Besides, It can be noticed that in term of BER, FER evaluations, our works outperform Bit-level Soft (BLS) RLL based on Reed-Solomon(RS) code solutions at code-rates (15/11), (15/7) and (15/3), which are first mentioned in \cite{Wang2}. Furthermore, BER, FER performances of our non-RLL solution also outperform our previous work which is based on soft-decoding of 4B6B and Polar code\cite{Dung}. Besides, we have selected other typical BER performances which are introduced in related works \cite{Sunghwan,Xuanxuan,Wang1} to make comparisons with our proposed receiver (Fig.\ref{fig6}).

\subsection{FPGA and ASIC Implementation of proposed VLC receiver}
Tab.\ref{table_3} and Tab.\ref{table_4} summarize the FPGA and ASIC synthesis results of our non-RLL VLC receiver. The proposed hardware architecture is defined by Verilog HDL and be verified by ModelSim. FPGA synthesis process is conducted by Altera Quartus II, while ASIC synthesis is done by Synopsys' Design Compiler. The selected technology library is VDEC Rohm 180nm. It can be inferred from Tab.\ref{table_3} that the Polar decoder is the largest block of our VLC decoder. In particular, Polar decoder takes 86\% logic resource of the whole receiver. The frequency of the proposed VLC receiver is also restricted at 25 Mhz (maximum frequency is 29.31 Mhz) due to the non-registered decoding architecture of successive cancellation (SC) Polar decoder. However, the max operation frequency can be increased by adding more buffering registers in the processing elements (PEs) network of Polar decoder with extra complexity and latency. We also estimate the throughout, energy-per-bit and hardware efficiency of the proposed receiver by Eq.(\ref{eq:Throughput})(\ref{eq:Energy})(\ref{eq:Hardware}), and results are presented in Tab.\ref{table_4}.

\begin{equation} \label{eq:Throughput}
\text{Thoughput [b/s]} = \dfrac{\text{$N$ [b]}}{\textrm{$D_N$ [sec]}} 
\end{equation}

\begin{equation} \label{eq:Energy}
\text{Energy-per-bit [J/b]} = \dfrac{\text{Power [W]}}{\text{Thoughput [b/s]}} 
\end{equation}

\begin{equation} \label{eq:Hardware}
\text{Hardware Efficiency [b/s/$m^2$]} = \dfrac{\text{Throughput [b/s]}}{\text{Area [$m^2$]}}
\end{equation}

\begin{table}[!t]
\renewcommand{\arraystretch}{1.3}
\caption{FPGA synthesys results of our VLC receiver}
\label{table_3}
\centering
\begin{tabular}{c | c | c }
\hline
\bfseries 	 &\bfseries Receiver  &\bfseries Polar Decoder  \\
\hline 				 \hline
Device     	 		 & Cyclone IV  		  &  	Cyclone IV  \\\hline
Model	     		 & 1200mV 0C   		  &     1200mV 0C   \\\hline
Fmax        		 & 29.31         	  &      28.87	    \\\hline
LE/LUT	    		 & 12134/114480 (10.6\%) &   10455/114480 (9\%)		\\\hline
Registers  			 & 3109  			  &     1544				\\\hline
Memory bits 		 & 1152  			  &       0  		\\\hline
\hline
\end{tabular}
\end{table}

\begin{table}[!t]
\renewcommand{\arraystretch}{1.3}
\caption{ASIC synthesis results of our VLC receiver}
\label{table_4}
\centering
\begin{tabular}{c | c}
\hline
\bfseries 	&\bfseries Receiver\\
\hline \hline
Technology [$nm$]         & 180       \\\hline
Voltage    [$V$]	       & 1.8       \\\hline
Area      [$\mu m^2$]  & 573724.56 \\\hline
Frequency [$Mhz$]            & 25        \\\hline
Power     [$mW$]          & 3.5022    \\\hline
Throughput [$Mb/s$]        & 16.58        \\\hline
Energy-per-bit [$pJ/b$]        & 211.2   \\\hline
Hardware Efficiency [$Mb/s/mm^2$]  & 28.75  \\\hline
Latency [$clock$]		    		    & 386	 \\\hline
\hline
\end{tabular}
\end{table}

\section{Conclusion}
We have proposed a non-RLL DC-balance solution consisting of a pre-scrambler based on a simple generating polynomial combined with a Polar encoder. The proposed method has a centralized bit probability distribution, with approximately equal numbers of zero and one bits. Moreover, the maximum run-length of bit-0 is reduced remarkably when a pre-scrambler is applied with a Polar encoder. Therefore, DC-balance can be maintained even with the short data frames of VLC-based beacon systems. Moreover, the non-RLL nature of the proposal reduces the complexity of the VLC receiver with great improvements on information code-rate. Besides, we also introduced a 3-bit soft-decision filter which enables soft decoding of polar decoder can be implemented in real VLC receiver prototypes to enhance the error-correction performances. Also, BER and FER performances of the proposed receiver are evaluated and some discussions on them are given. Finally, we have introduced a novel hardware architecture for the proposed non-RLL VLC receiver with FPGA and ASIC synthesis results are given in details.

\section*{Acknowledgment}
This work  was supported by JSPS KAKENHI Grant Number JP16K18105.



%

\end{document}